\documentclass[10pt,a4paper,twoside,final]{article}
\usepackage[switch]{lineno}

\usepackage[a4paper,
		top=2.5cm,
		bottom=1cm,
		textwidth=183mm,
		includefoot,
		foot=1cm,
		head=2cm]{geometry}

\usepackage{amsmath,amsfonts, amssymb, amsthm}

\usepackage{sidecap}

\usepackage{wasysym}
\usepackage{xcolor, framed}
\definecolor{shadecolor}{rgb}{0.9, 0.9, 0.9}

\usepackage[]{graphicx}

\usepackage{multicol, multirow}

\usepackage[font=small,labelfont=bf,sc]{caption}

\usepackage{enumerate, listings, siunitx}

\usepackage{mdframed}
\usepackage{calc}
\newlength{\mylength}
\newlength{\figurewidth}
\newsavebox\myboxA
\newsavebox\myboxB
\newlength\mylenA
\newcommand*\widebar[2][0.8]{%
    \sbox{\myboxA}{$\m@th#2$}%
    \setbox\myboxB\null
    \ht\myboxB=\ht\myboxA%
    \dp\myboxB=\dp\myboxA%
    \wd\myboxB=#1\wd\myboxA
    \sbox\myboxB{$\m@th\overline{\copy\myboxB}$}
    \setlength\mylenA{\the\wd\myboxA}
    \addtolength\mylenA{-\the\wd\myboxB}%
    \ifdim\wd\myboxB<\wd\myboxA%
       \rlap{\hskip 0.5\mylenA\usebox\myboxB}{\usebox\myboxA}%
    \else
        \hskip -0.5\mylenA\rlap{\usebox\myboxA}{\hskip 0.5\mylenA\usebox\myboxB}%
    \fi}

\usepackage[numbers,comma,square]{natbib}
\renewcommand{\cite}{\citep}
\renewcommand{\citet}{\citep}

\newcommand{\opener}{}

\title{25 years of criticality in neuroscience -- established results, open controversies, novel concepts}

\author{J. Wilting$^1$ and V. Priesemann$^{1,2}$ \\ {\small $^1$Max-Planck-Institute for Dynamics and Self-Organization, G\"ottingen} \\ {\small $^2$Bernstein-Center for Computational Neuroscience, G\"ottingen}}
\date{ \normalsize Preprint, \today}

\begin{document}
\twocolumn[
  \begin{@twocolumnfalse}

   {\Huge  \noindent 25 years of criticality in neuroscience -- established results, open controversies, novel concepts} \\
   
   {\large \noindent J. Wilting$^1$ \& V. Priesemann$^{1,2,*}$ }\\

   $^1$Max-Planck-Institute for Dynamics and Self-Organization, G\"ottingen, Germany; $^2$Bernstein-Center for Computational Neuroscience, G\"ottingen, Germany\\
   $^*$ viola.priesemann@ds.mpg.de

\vspace*{0.1cm}

\begin{shaded}
\noindent
\textbf{
\noindent
Twenty-five years ago, Dunkelmann and Radons (1994) proposed that neural networks should self-organize to a critical state. In models, criticality offers a number of computational advantages. Thus this hypothesis, and in particular the experimental work by Beggs and Plenz (2003), has triggered an avalanche of research, with thousands of studies referring to it. Nonetheless, experimental results are still contradictory. How is it possible, that a hypothesis has attracted active research for decades, but nonetheless remains controversial?
We discuss the experimental and conceptual controversy, and then present a parsimonious solution that (i) unifies the contradictory experimental results, (ii) avoids disadvantages of a critical state, and (iii) enables rapid, adaptive tuning of network properties to task requirements.
}
\end{shaded}

\vspace*{0.1cm}
  \end{@twocolumnfalse}
]

\subsection*{Highlights}
\begin{itemize}
\item The criticality hypothesis has received major attention in the past 25 years
\item We revise and discuss the experimental and conceptual controversies
\item We propose that cortical dynamics is reverberating, subcritical
\item In vitro neuronal networks indeed self-organize to a critical state
\item Thereby, many of the controversies can be resolved
\end{itemize}


\subsection*{Introduction}

\begin{SCfigure*}
\includegraphics[width=\columnwidth]{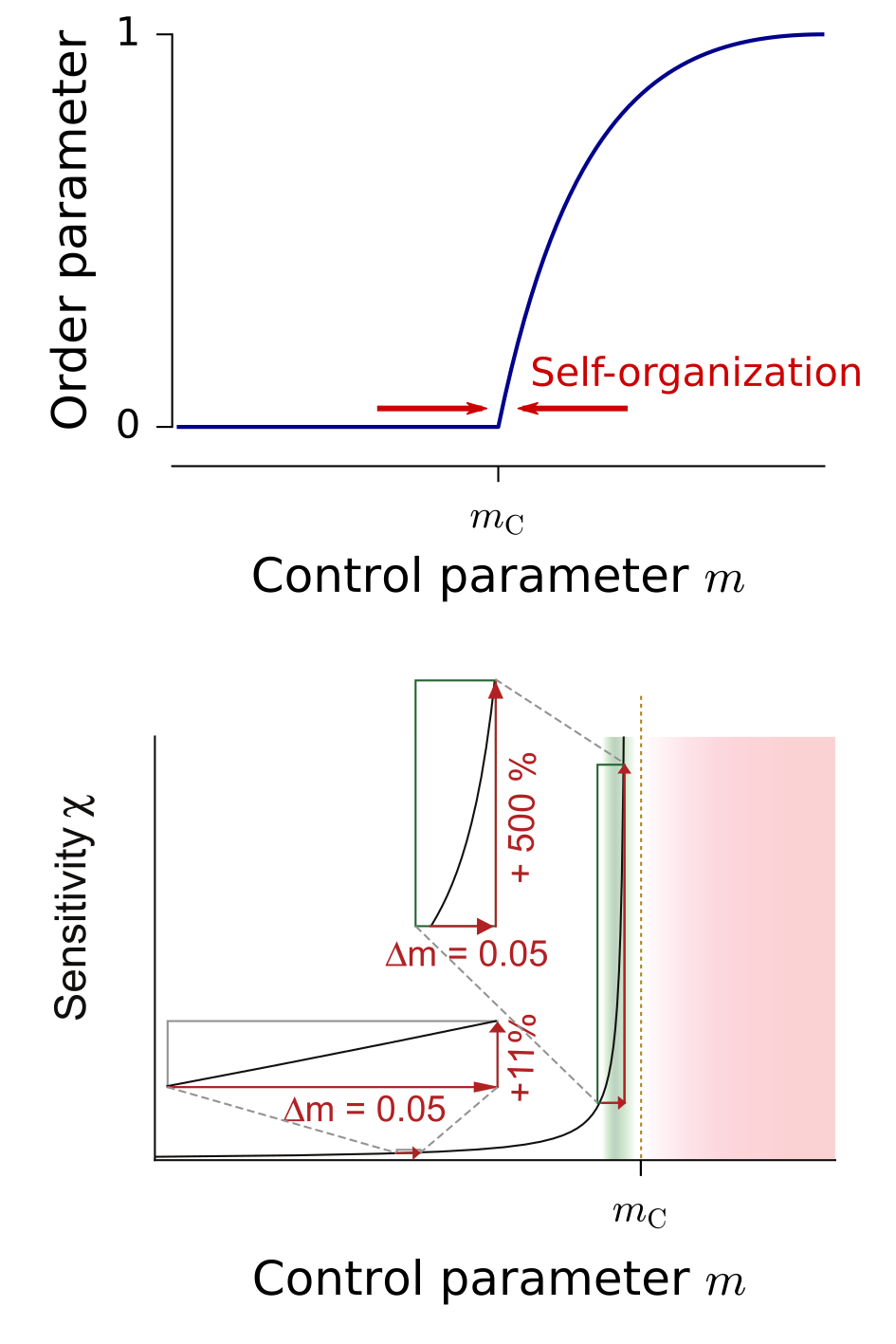}
\caption{\textbf{Criticality.} \textbf{a.} In general, if the dynamics of a network can be adjusted by  a control parameter $m$ (or $\sigma$), a critical point $m_\mathrm{C}$ marks the transition between distinct dynamical regimes.
One can define an order parameter, which is zero in one regime, but non-zero in the other.
For this 2nd order phase transition, the change of the order parameter is continuous, but not smooth at $m_\mathrm{C}$.
For avalanche criticality, this order parameter is the probability, that an avalanche lasts forever, and thereby marks the transition from stable ($m < m_\mathrm{C}$) to instable ($m > m_\mathrm{C}$) dynamics.
Other phase transitions could be between ordered and chaotic, oscillatory and non-osciallatory, or asynchronous and synchronized dynamics.
If the critical state is an attractor state, i.e. the control parameter is driven back to the critical point (red arrows), this is called self-organized criticality.
Certain plasticity rules like homeostastic plasticity or synaptic depletion can implement self-organized criticality \citep[e.g.][]{Levina2007,Levina2009a,Zierenberg2018}, but could also promote adaptation to different states \citep{Zierenberg2018}.
\textbf{b.} Many network properties, like the sensitivity, variance of responses, correlation length, and autocorrelation time, diverge at the critical point. Therefore, networks poised in a \textit{reverberating regime} (green) can adjust these network properties by small changes in the control parameter, much more rapidly than networks poised farther away from criticality (insets).}
\label{box1}
\end{SCfigure*}

\opener{Twenty-five years ago, \citet{Dunkelmann1994} proposed that neural networks might self-organize to a critical state.}
This critical state marks the transition between stable and unstable dynamics (Box \ref{box1}): On average, the network conserves the number of spikes, thus every spike in one neuron on average causes one spike in all its postsynaptic neurons~\citep{Miranda1991,Jensen1998,Pruessner2012,Munoz2018,Roli2018}.
The critical state is characterized by spatio-temporal cascades of activity, called avalanches, that typically are very small, but some span the entire network.

\opener{In the past twenty-five years, the criticality hypothesis, and in particular the experimental work by John Beggs and Dietmar Plenz (2003), has triggered an avalanche of research, with thousands of studies referring to it.}
Nonetheless, experimental results are still contradictory.
How is it possible, that a hypothesis is at the same time  so attractive and fascinating that it has prevailed for more than two decades, but also sparked heated debates and still remains controversial?

\opener{The hypothesis that the brain operated at a critical point is attractive for two reasons.}
On a conceptual level, criticality has been shown to maximize a number of properties that are considered favourable for computation.
On an experimental level, there is considerable evidence in support of the criticality hypothesis.
\opener{However, both of these points are discussed controversially.}
On the conceptual level, maximization of certain properties is unlikely to explain cortical function, and it is frequently neglected that criticality also maximizes properties that are likely adversarial to cortical function.
On the experimental level, assessing criticality is more intricate than first thought, undermining the significance of the accumulated evidence.
In the following, we discuss these two points in detail.

\subsection*{Conceptual appeal and controversies}

\opener{In models of neural networks, criticality maximizes a number of properties considered favourable for computation~\cite{Beggs2008,Shew2013}.}
Tuning models towards critical phase transitions has been shown to maximize the number of metastable states~\cite{Haldeman2005}, the dynamic range~\cite{Kinouchi2006,Gautam2015}, information transmission in terms of mutual information~\cite{Tanaka2009,Shriki2013,Shriki2016}, active information storage~\cite{Boedecker2012}, and computational power in terms of input-output mappings~\cite{Maass2002,Bertschinger2004,Latham2004}.
Self-organized critical networks can also efficiently implement non-convex optimization \citep{Hoffmann2018}.
These functional benefits made criticality an attractive target state for self-organization of cortical networks.

\opener{However, criticality also maximizes further aspects, which are likely negative for function.}
For example, the variability of network responses diverges at the critical point, which comes at the cost of reduced specificity \cite{Gollo2017} and reliability \cite{Wilting2018a}.
Likewise, criticality is accompanied by ``critical slowing down''~\citep{Scheffer2012}, which means that systems might take overly long to finish a computation before being ready for new stimuli.

\opener{Instead of \textit{maximizing} particular network properties, \textit{sufficient} performance for a given task is a more likely design principle for network function \citep{Boedecker2012}.}
This sufficient performance is most likely not achieved by maximizing one particular property alone, but rather by balancing the competing, positive, and negative  aspects.
This balance may represent, e.g., a balance between sensitivity and specificity~\cite{Gollo2017}, between quality of representation and integration time~\cite{Shriki2016}, or between stimulus detection and discrimination \citep{Tomen2014,Clawson2017}.
Importantly, this optimal balance may not be achieved at criticality.

\subsection*{Experimental evidence and challenges}

\begin{figure*}
\includegraphics[width=170mm]{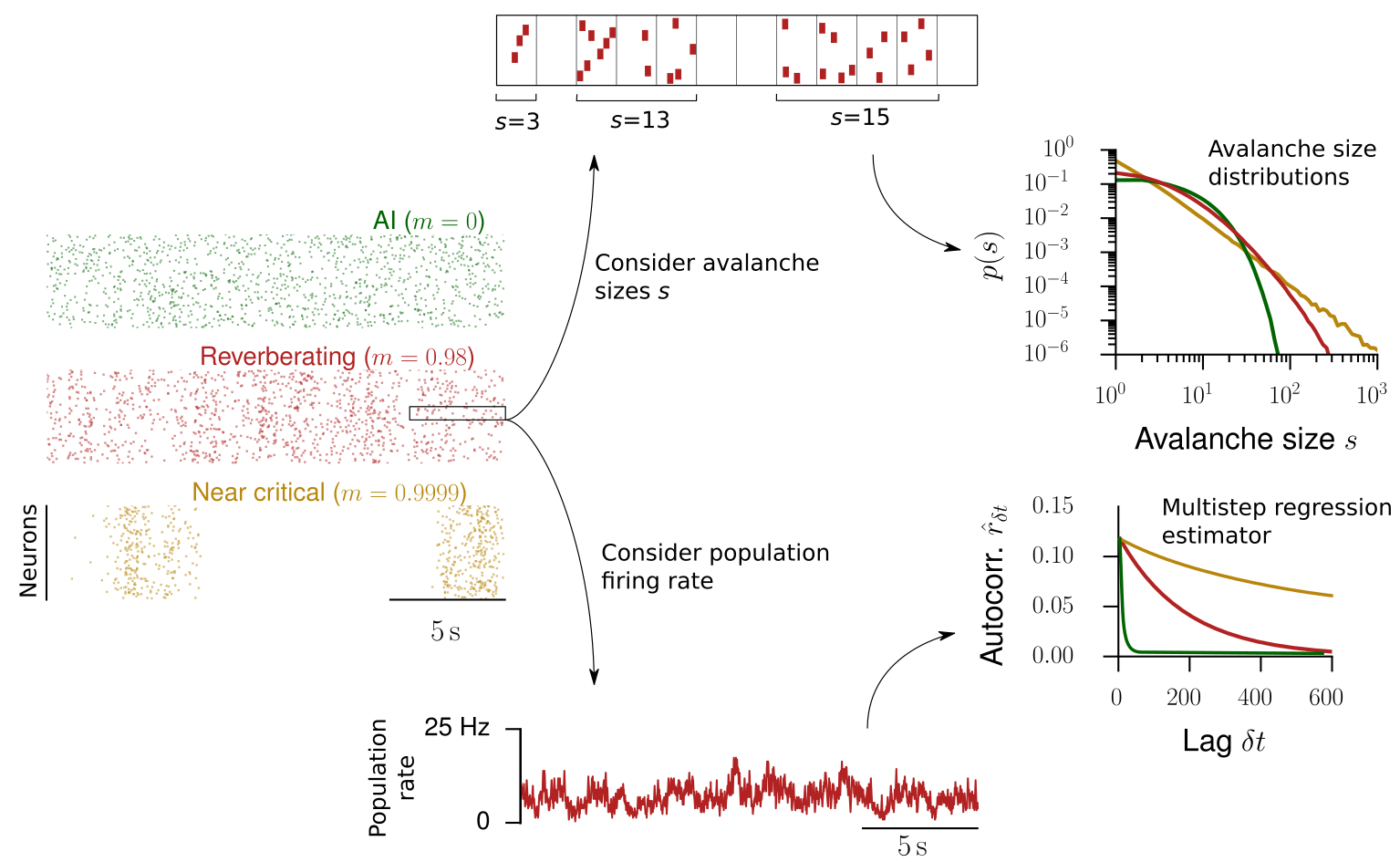}
\caption{\textbf{Assessing criticality in experiments.} 
\textbf{Left.} 
Raster plots of the population activity for three different dynamics, asynchronous and irregular (AI), reverberating, and near critical.
\textbf{Upper part.} A subset of all neurons is recorded, and avalanches are extracted using temporal binning. Empty bins (pauses) separate subsequent avalanches. The avalanche size $s$ is the total number of spikes in a sequence of non-empty bins. In critical systems, the avalanche size distribution $p(s)$ follows an (approximate) power law if the system is critical, even if only a random subset of neurons was sampled~\citep{Levina2017}, whereas for Poisson activity it is approximately exponential~\citep{Priesemann2018}.
\textbf{Lower part.} The ongoing activity considers the spikes in each bin, without isolating avalanches. From this ongoing activity, the multistep regression estimator \citep{Wilting2018} allows to infer the distance to criticality from the autocorrelation of the ongoing population activity.
}
\label{box2}
\end{figure*}

\opener{Assessing criticality in experiments traditionally relies on identifying power-law distributions of avalanche sizes.}
Avalanches are spatio-temporal cascades of activity, whose sizes are expected to follow a power-law distribution \cite{Bak1987} if networks were critical (see Box \ref{box2}).
The slope of this distribution depends on the universality class of the underlying dynamics~\cite{Jensen1998,Pruessner2012} and is, e.g., expected to be $-3/2$ for critical branching processes~\cite{Harris1963}.

\opener{Power law distributions of avalanche sizes were indeed found in many experiments.}
Following the influential study by \citet{Beggs2003}, who identified power-law distributed avalanches in the LFPs of cultured slices, subsequent studies \textit{in vitro} found similar distributions in LFPs and spiking activity of cultured dissociated neurons or cortical slices~\citep{Mazzoni2007,Pasquale2008,Tetzlaff2010,Friedman2012}.
In living animals, power-law distributions were found in LFPs of macaques~\citep{Priesemann2009}, rhesus monkeys~\citep{Petermann2009,Yu2017}, and cats~\citep{Hahn2010,Hahn2017}, in two-photon microscopy in mice \citep{Sharp2014,Bellay2015}, in light-sheet microscopy of whole-brain GCaMP activity in zebrafish larvae~\citep{Ponce-Alvarez2018}, in LFPs from intracranial depth electrodes in humans~\citep{Priesemann2013}, in MEG and EEG of adult humans~\citep{Linkenkaer-Hansen2001,Shriki2013,Palva2013} and preterm babies~\citep{Iyer2015}, and in human BOLD signals~\citep{Tagliazucchi2012}.

\opener{Avalanche size distribution of  \textit{in vivo} spiking activity often differ from power-laws.}
While most evidence for criticality \textit{in vivo}, in particular the characteristic power-law distributions, has been obtained from \textit{coarse} measures of neural activity (LFP, EEG, BOLD); see \cite{Priesemann2014} and references therein), avalanches reconstructed from spiking activity in awake animals typically did not display power laws \cite{Priesemann2014, Ribeiro2010, Bedard2006, Pasquale2008, Hahn2010,Tetzlaff2010}.
These deviations from power-laws, were attributed to subsampling effects~\citep{Priesemann2009,Ribeiro2010}, i.e. the limitation that only a small number of neurons can be recorded simultaneously with millisecond precision \cite{Priesemann2009,Priesemann2013,Girardi-Schappo2013,Priesemann2014,Ribeiro2014,Levina2017,Wilting2018}.
However, power laws should retain approximate power-law characteristics under random subsampling~\cite{Levina2017}.
Therefore we argue that these deviations from power-laws actually reflect deviations from critical dynamics, as outlined below.

\opener{The extraction of avalanches from neural recordings can be ambiguous.}
Any two subsequent avalanches are separated by pauses in the activity, i.e. bins without spikes (see Box \ref{box2}).
These empty bins define the end of one avalanche and the start of a new one.
However, as neural activity does not show a separation of timescales (i.e. clear pauses) that allows for an unambiguous separation of avalanches, because the brain (or a reasonably sized part of the brain, e.g. a cortical column) is certainly never silent for several milliseconds. 
Instead, avalanches are continuously initialized, merge and split up, presenting a melange that cannot be separated into the individual cascades from spike recordings alone. 
Hence, for analysis purpose, empty bins are only found if (i) the bin size tends towards zero, (ii) the system is heavily subsampled, or (iii) thresholding is applied to the activity~\cite{DelPapa2017,Poil2012,Priesemann2014}.
This dependence on the experimental setting and analysis parameters renders the definition of avalanches ambiguous. As a consequence, the resulting avalanche distribution depends on the choice of the bin size, threshold, and the number of analyzed neurons or channels.

\opener{Power-law distributions may also arise in the absence of criticality.}
\citet{Priesemann2018} showed that time-varying network input can give rise to power-law distributed avalanches even though the underlying network did not self-organize to criticality.
Furthermore, power-law distributions may arise simply because of logarithmic representations or thresholding~\cite{Touboul2010}.
Hence, power-law statistics alone are not necessarily indicative of self-organized criticality.

\opener{Recent advances in the study of subsampled systems overcome many of the ambiguities of avalanche size analysis}
A novel estimator that relies on ongoing neural activity can precisely quantify the distance from criticality on a continuous spectrum \cite{Wilting2018}, and is easy to apply to data (Box \ref{box2}): It requires only a few tens of multi-unit spike recordings, is subsampling-invariant, and does not require a separation of timescales, or spike sorting.
It can reliably infer the distance from criticality, and is thus is a valuable tool to investigate cortical network dynamics, and the changes thereof.

\subsection*{Reverberating dynamics}

\opener{This novel estimator allowed to identify that cortical dynamics is not critical, but reverberating.}
We applied the estimator to \textit{in vivo} spike recordings and identified a reverberating regime ($0.94 < m < 0.998$), 
consistently across brain areas, species, and tasks \cite{Wilting2018, Wilting2019}.
This reverberating regime has also been found by a complementary approach by \citet{Dahmen2016}, who inferred it from the distributions of spike covariances.

\opener{This reverberating regime may resolve many of the conceptual controversies \citep{Wilting2018a,Wilting2019}.}
First, instead of solely maximizing singular properties, the reverberating regime can balance competing requirements, e.g. sensitivity vs. specificity \cite{Gollo2017}, quality of representation vs. integration time~\cite{Shriki2016}, or stimulus detection vs. discrimination \citep{Tomen2014,Clawson2017}.
Second, the reverberating regime allows flexible adaption to task requirements, because small parameter changes can induce rapid, string changes of computational properties, which we called dynamic adaptive computation \cite{Wilting2018a} (see Box \ref{box1}).
Third, it allows amplification of small stimuli, while keeping a safety margin from supercriticality, which has been associated with instability.
Fourth, the reverberating regime offers the substrate to tune short-term memory, as information about the input is maintained for well-defined time-spans ranging from a few hundred milliseconds to a few seconds, which has been exploited in echo state networks \cite{Maass2002,Jaeger2007,Boedecker2012}.
Last, the reverberating regime offers a loop hole from the platonic idea of criticality.
In contrast to the view of ``as close to critical as possible'', which still holds criticality as the ideal target \cite{Williams-Garcia2014,Martinello2017}, cortical networks might self-organize to different points in the reverberating regime, and thereby gain flexibility and the ability to balance competing requirements.

\opener{The reverberating regime may also resolve experimental controversies.}
Considering experiments in humans or awake animals, most power-law distributions, as evidence for criticality, have been found in coarse measures like population spikes, LFP, EEG or MEG signals. For all these signals, the electrodes may sample from overlapping populations. This enhances the \textit{observed} correlations between electrodes, and can lead to power-laws, even if the underlying population is not critical but in a reverberating regime \citep{Neto2017}.
In contrast,  in spike recordings \textit{in vivo} power laws were typically not observed \citep{Priesemann2014,Ribeiro2010,Bedard2006,Dehghani2012}. A parsimonious explanation is that \textit{in vivo}, cortical dynamics is not critical, but resembles a reverberating regime.
This does not contradict the evidence for criticality --- including spiking activity --- in \textit{in vitro} setups.
The emergence of very different states, a reverberating state \textit{in vivo} and a critical state \textit{in vitro}, can be explained by differences in topology \citep{Kossio2018}, or the lack of external input characterizing isolated \textit{in vitro} networks \citet{Zierenberg2018}.

\subsection*{Open topics}

\opener{It is unclear how different concepts of criticality agree.}
The term criticality is not strictly defined and used for multiple concepts \citep{Munoz2018}.
Besides the avalanche criticality discussed in this review, which is a transition between stability and instability, other concepts of dynamical criticality exist.
First, the transition could be between ordered and chaotic, called the edge of chaos \citep{Bertschinger2004,Boedecker2012}.
Second, it could be between non-oscillating and oscillating, called the edge of synchrony \citep{Brunel2000,Markram2015,Palmigiano2017}.
Although these three concepts share some features like avalanche size distributions \citep{DiSanto2018} and critical slowing down~\citep{Scheffer2012}, a definite justification for their interchangable use in the literature remains desirable.
Apart from dynamical criticality, neural networks might exhibit statistical criticality, identified by diverging specific heat in maximum entropy models \citep{Mora2011,Tkacik2013,Tkacik2015,Humplik2017,Savin2017,Nonnenmacher2017}.
How statistical and dynamical criticality precisely relate to each other is a topic of open research.

\opener{Dynamic adaptive computation allows experimental predictions to be tested in the future.}
A hierarchy of distances to criticality and the corresponding network timescales has been observed across visual and sensory pathways \cite{Murray2014a} and may represent a specialization to network function \cite{Hasson2008,Honey2012,Chen2015,Chaudhuri2015}.
The concept of dynamic adaptive computation predicts that a modulation of this hierarchical backbone can be observed on various timescales, and depends on vigilance state and task.
Indeed, there is evidence that the power-law nature of avalanche size distributions changes with cognitive states, for example across the sleep-wake cycle \citep{Ribeiro2010,Priesemann2013,Meisel2017a,Meisel2017b} or under changes of consciousness \citep{Tagliazucchi2016,Bellay2015,Fagerholm2016,Fagerholm2018,Fekete2018,Lee2019}.
Markers of criticality have been shown to transiently change depending on the behavioral state \citep{Hahn2017}, attention \citep{Fagerholm2015}, and stimuli \cite{Arviv2015,Yu2017}. 
These studies provided evidence that the brain may not be self-organized critical, but is capable of self-organization to different states.
Precisely linking the changes in the reverberating regime to the specific task at hand remains a challenge for future experiments.

\paragraph{Acknowledgments}
JW and VP received financial support from the Max Planck Society.
VP: ADD: Acknowledgement to group for discussion!

\paragraph{Conflict of interests}
The authors declare that there was no conflict of interests.

\subsection*{Annotated references}
\textbf{**} \citet{Munoz2018} extensively reviews different concepts of criticality in biological systems, including neural networks.\\
\textbf{**} \citet{Beggs2003} showed the first experimental evidence for criticality in neural systems.\\
\textbf{*} \citet{Wilting2018} introduced a novel method to gauge criticality beyond avalanche size distributions.\\
\textbf{*} \citet{Levina2009a} constructed the first model to self-organize to criticality over a large parameter range.\\
\textbf{*} \citet{Boedecker2012} shows that different characterizations of information processing are maximized at criticality.\\
\textbf{*} \citet{Barnett2013} show that transfer entropy is not maximized at criticality, but in a subcritical regime.\\
\textbf{*} \citet{Chaudhuri2015} used a model poised close to a phase transition to construct a model for hierarchical processing.\\
\textbf{*} \citet{Zierenberg2018} demonstrated that homeostatic plasticity can mediate self-organization to diverse dynamical states, depending on the network input.\\
\textbf{*} \citet{Clawson2017} explicitly demonstrated a trade-off between different computational properties at criticality, at the example of stimulus detection vs. discrimination.\\
\textbf{*} \citet{Savin2017} is a recent review of thermodynamic criticality in neuronal networks.

\begin{scriptsize}

\begin{thebibliography}{10}

\bibitem{Dunkelmann1994}
S.~Dunkelmann and G.~Radons.
\newblock {Neural Networks and Abelian Sandpile Models of Self-Organized
  Criticality}.
\newblock In M.~Marinaro and P.~G. Morasso, editors, {\em ICANN '94
  (Proceedings of the International Conference on Artificial Neural Networks)},
  page 867. Springer-Verlag, 1994.

\bibitem{Miranda1991}
E.N. Miranda and H.J. Herrmann.
\newblock {Self-organized criticality with disorder and frustration}.
\newblock {\em Physica A: Statistical Mechanics and its Applications},
  175(3):339--344, jul 1991.

\bibitem{Jensen1998}
Henrik~Jeldtoft Jensen.
\newblock {\em {Self-Organized Criticality --- Emergent Complex Behavior in
  Physical and Biological Systems}}.
\newblock Cambridge University Press, Cambridge, 1998.

\bibitem{Pruessner2012}
Gunnar Pruessner.
\newblock {\em {Self-organised criticality: theory, models and
  characterisation}}.
\newblock Cambridge University Press, 2012.

\bibitem{Munoz2018}
Miguel~A. Mu{\~{n}}oz.
\newblock {Colloquium : Criticality and dynamical scaling in living systems}.
\newblock {\em Reviews of Modern Physics}, 90(3):031001, jul 2018.

\bibitem{Roli2018}
Andrea Roli, Marco Villani, Alessandro Filisetti, and Roberto Serra.
\newblock {Dynamical Criticality: Overview and Open Questions}.
\newblock {\em Journal of Systems Science and Complexity}, 31(3):647--663,
  2018.

\bibitem{Levina2007}
a.~Levina, J.~M. Herrmann, and T.~Geisel.
\newblock {Dynamical synapses causing self-organized criticality in neural
  networks}.
\newblock {\em Nature Physics}, 3(12):857--860, nov 2007.

\bibitem{Levina2009a}
Anna Levina, J.~Herrmann, and Theo Geisel.
\newblock {Phase Transitions towards Criticality in a Neural System with
  Adaptive Interactions}.
\newblock {\em Physical Review Letters}, 102(11):118110, mar 2009.

\bibitem{Zierenberg2018}
Johannes Zierenberg, Jens Wilting, and Viola Priesemann.
\newblock {Homeostatic Plasticity and External Input Shape Neural Network
  Dynamics}.
\newblock {\em Physical Review X}, 8(3):031018, 2018.

\bibitem{Beggs2008}
John~M Beggs.
\newblock {The criticality hypothesis: how local cortical networks might
  optimize information processing}.
\newblock {\em Philosophical Transactions of the Royal Society of London A},
  366(1864):329--343, 2008.

\bibitem{Shew2013}
Woodrow~L Shew and Dietmar Plenz.
\newblock {The functional benefits of criticality in the cortex.}
\newblock {\em The Neuroscientist : a review journal bringing neurobiology,
  neurology and psychiatry}, 19(1):88--100, feb 2013.

\bibitem{Haldeman2005}
Clayton Haldeman and John Beggs.
\newblock {Critical Branching Captures Activity in Living Neural Networks and
  Maximizes the Number of Metastable States}.
\newblock {\em Physical Review Letters}, 94(5):058101, feb 2005.

\bibitem{Kinouchi2006}
Osame Kinouchi and Mauro Copelli.
\newblock {Optimal dynamical range of excitable networks at criticality}.
\newblock {\em Nature Physics}, 2(5):348--351, apr 2006.

\bibitem{Gautam2015}
Shree~Hari Gautam, Thanh~T. Hoang, Kylie McClanahan, Stephen~K. Grady, and
  Woodrow~L. Shew.
\newblock {Maximizing Sensory Dynamic Range by Tuning the Cortical State to
  Criticality}.
\newblock {\em PLoS Computational Biology}, 11(12):1--15, 2015.

\bibitem{Tanaka2009}
Takuma Tanaka, Takeshi Kaneko, and Toshio Aoyagi.
\newblock {Recurrent infomax generates cell assemblies, neuronal avalanches,
  and simple cell-like selectivity}.
\newblock {\em Neural Computation}, 21(4):1038--1067, 2009.

\bibitem{Shriki2013}
Oren Shriki, Jeff Alstott, Frederick Carver, Tom Holroyd, Richard N~a Henson,
  Marie~L Smith, Richard Coppola, Edward Bullmore, and Dietmar Plenz.
\newblock {Neuronal avalanches in the resting MEG of the human brain.}
\newblock {\em The Journal of neuroscience : the official journal of the
  Society for Neuroscience}, 33(16):7079--90, apr 2013.

\bibitem{Shriki2016}
Oren Shriki and Dovi Yellin.
\newblock {Optimal Information Representation and Criticality in an Adaptive
  Sensory Recurrent Neuronal Network}.
\newblock {\em PLoS Computational Biology}, 12(2):1--19, 2016.

\bibitem{Boedecker2012}
Joschka Boedecker, Oliver Obst, Joseph~T Lizier, N~Michael Mayer, and Minoru
  Asada.
\newblock {Information processing in echo state networks at the edge of chaos}.
\newblock {\em Theory in Biosciences}, 131(3):205--213, sep 2012.

\bibitem{Maass2002}
Wolfgang Maass, Thomas Natschl{\"{a}}ger, and Henry Markram.
\newblock {Real-Time Computing Without Stable States: A New Framework for
  Neural Computation Based on Perturbations}.
\newblock {\em Neural Computation}, 14(11):2531--2560, nov 2002.

\bibitem{Bertschinger2004}
Nils Bertschinger and Thomas Natschl{\"{a}}ger.
\newblock {Real-Time Computation at the Edge of Chaos in Recurrent Neural
  Networks}.
\newblock {\em Neural Computation}, 16(7):1413--1436, jul 2004.

\bibitem{Latham2004}
Peter~E. Latham and Sheila Nirenberg.
\newblock {Computing and stability in cortical networks}.
\newblock {\em Neural Computation}, 16(7):1385--1412, 2004.

\bibitem{Hoffmann2018}
Heiko Hoffmann and David~W. Payton.
\newblock {Optimization by Self-Organized Criticality}.
\newblock {\em Scientific Reports}, 8(1):1--9, 2018.

\bibitem{Gollo2017}
Leonardo~L Gollo.
\newblock {Coexistence of critical sensitivity and subcritical specificity can
  yield optimal population coding}.
\newblock {\em Journal of The Royal Society Interface}, 14(134):20170207, sep
  2017.

\bibitem{Wilting2018a}
Jens Wilting, Jonas Dehning, Joao {Pinheiro Neto}, Lucas Rudelt, Michael
  Wibral, Johannes Zierenberg, and Viola Priesemann.
\newblock {Operating in a Reverberating Regime Enables Rapid Tuning of Network
  States to Task Requirements}.
\newblock {\em Frontiers in Systems Neuroscience}, 12(November):55, 2018.

\bibitem{Scheffer2012}
Marten Scheffer, Stephen~R Carpenter, Timothy~M Lenton, Jordi Bascompte,
  William Brock, Vasilis Dakos, Johan van~de Koppel, Ingrid~A van~de Leemput,
  Simon~A Levin, Egbert~H van Nes, Mercedes Pascual, and John Vandermeer.
\newblock {Anticipating critical transitions}.
\newblock {\em Science}, 338:344--348, 2012.

\bibitem{Tomen2014}
Nergis Tomen, David Rotermund, and Udo Ernst.
\newblock {Marginally subcritical dynamics explain enhanced stimulus
  discriminability under attention}.
\newblock {\em Frontiers in Systems Neuroscience}, 8(August):1--15, 2014.

\bibitem{Clawson2017}
Wesley~P. Clawson, Nathaniel~C. Wright, Ralf Wessel, and Woodrow~L. Shew.
\newblock {Adaptation towards scale-free dynamics improves cortical stimulus
  discrimination at the cost of reduced detection}.
\newblock {\em PLoS Computational Biology}, 13(5):1--21, 2017.

\bibitem{Levina2017}
A.~Levina and V.~Priesemann.
\newblock {Subsampling scaling}.
\newblock {\em Nature Communications}, 8(May):15140, may 2017.

\bibitem{Priesemann2018}
Viola Priesemann and Oren Shriki.
\newblock {Can a time varying external drive give rise to apparent criticality
  in neural systems?}
\newblock {\em PLOS Computational Biology}, 14(5):e1006081, 2018.

\bibitem{Wilting2018}
Jens Wilting and Viola Priesemann.
\newblock {Inferring collective dynamical states from widely unobserved
  systems}.
\newblock {\em Nature Communications}, 9(1):2325, 2018.

\bibitem{Bak1987}
Per Bak, Chao Tang, and Kurt Wiesenfeld.
\newblock {Self-organized criticality: An explanation of the 1/f noise}.
\newblock {\em Physical Review Letters}, 59(4):381--384, jul 1987.

\bibitem{Harris1963}
Theodore~E. Harris.
\newblock {\em {The Theory of Branching Processes}}.
\newblock Springer Berlin, 1963.

\bibitem{Beggs2003}
John~M Beggs and Dietmar Plenz.
\newblock {Neuronal Avalanches in Neocortical Circuits}.
\newblock {\em The Journal of Neuroscience}, 23(35):11167--11177, dec 2003.

\bibitem{Mazzoni2007}
Alberto Mazzoni, Fr{\'{e}}d{\'{e}}ric~D. Broccard, Elizabeth Garcia-Perez,
  Paolo Bonifazi, Maria~Elisabetta Ruaro, and Vincent Torre.
\newblock {On the dynamics of the spontaneous activity in neuronal networks}.
\newblock {\em PLoS ONE}, 2(5), 2007.

\bibitem{Pasquale2008}
V~Pasquale, P~Massobrio, L~L Bologna, M~Chiappalone, and S~Martinoia.
\newblock {Self-organization and neuronal avalanches in networks of dissociated
  cortical neurons.}
\newblock {\em Neuroscience}, 153(4):1354--69, jun 2008.

\bibitem{Tetzlaff2010}
Christian Tetzlaff, Samora Okujeni, Ulrich Egert, Florentin
  W{\"{o}}rg{\"{o}}tter, and Markus Butz.
\newblock {Self-organized criticality in developing neuronal networks.}
\newblock {\em PLoS computational biology}, 6(12):e1001013, jan 2010.

\bibitem{Friedman2012}
Nir Friedman, Shinya Ito, Braden a.~W. Brinkman, Masanori Shimono, R.~E.~Lee
  DeVille, Karin~a. Dahmen, John~M. Beggs, and Thomas~C. Butler.
\newblock {Universal Critical Dynamics in High Resolution Neuronal Avalanche
  Data}.
\newblock {\em Physical Review Letters}, 108(20):208102, may 2012.

\bibitem{Priesemann2009}
Viola Priesemann, Matthias H~J Munk, and Michael Wibral.
\newblock {Subsampling effects in neuronal avalanche distributions recorded in
  vivo.}
\newblock {\em BMC neuroscience}, 10:40, jan 2009.

\bibitem{Petermann2009}
Thomas Petermann, Tara~C Thiagarajan, Mikhail~a Lebedev, Miguel a~L Nicolelis,
  Dante~R Chialvo, and Dietmar Plenz.
\newblock {Spontaneous cortical activity in awake monkeys composed of neuronal
  avalanches.}
\newblock {\em Proceedings of the National Academy of Sciences of the United
  States of America}, 106(37):15921--6, sep 2009.

\bibitem{Yu2017}
Shan Yu, Tiago~L. Ribeiro, Christian Meisel, Samantha Chou, Andrew Mitz,
  Richard Saunders, and Dietmar Plenz.
\newblock {Maintained avalanche dynamics during task-induced changes of
  neuronal activity in nonhuman primates}.
\newblock {\em eLife}, 6:1--22, nov 2017.

\bibitem{Hahn2010}
Gerald Hahn, Thomas Petermann, Martha~N Havenith, Shan Yu, Wolf Singer, Dietmar
  Plenz, and Danko Nikolic.
\newblock {Neuronal avalanches in spontaneous activity in vivo.}
\newblock {\em Journal of neurophysiology}, 104(6):3312--22, dec 2010.

\bibitem{Hahn2017}
Gerald Hahn, Adrian Ponce-Alvarez, Cyril Monier, Giacomo Benvenuti, Arvind
  Kumar, Fr{\'{e}}d{\'{e}}ric Chavane, Gustavo Deco, and Yves Fr{\'{e}}gnac.
\newblock {Spontaneous cortical activity is transiently poised close to
  criticality}.
\newblock {\em PLoS Computational Biology}, 13(5):1--29, 2017.

\bibitem{Sharp2014}
D.~J. Sharp, G.~Scott, H.~Mutoh, E.~D. Fagerholm, T.~Knopfel, W.~L. Shew, and
  R.~Leech.
\newblock {Voltage Imaging of Waking Mouse Cortex Reveals Emergence of Critical
  Neuronal Dynamics}.
\newblock {\em Journal of Neuroscience}, 34(50):16611--16620, 2014.

\bibitem{Bellay2015}
Timothy Bellay, Andreas Klaus, Saurav Seshadri, and Dietmar Plenz.
\newblock {Irregular spiking of pyramidal neurons organizes as scale-invariant
  neuronal avalanches in the awake state}.
\newblock {\em eLife}, 4(JULY 2015):1--25, 2015.

\bibitem{Ponce-Alvarez2018}
Adri{\'{a}}n Ponce-Alvarez, Adrien Jouary, Martin Privat, Gustavo Deco, and
  Germ{\'{a}}n Sumbre.
\newblock {Whole-Brain Neuronal Activity Displays Crackling Noise Dynamics}.
\newblock {\em Neuron}, pages 1446--1459, 2018.

\bibitem{Priesemann2013}
Viola Priesemann, Mario Valderrama, Michael Wibral, and Michel {Le Van Quyen}.
\newblock {Neuronal avalanches differ from wakefulness to deep sleep--evidence
  from intracranial depth recordings in humans.}
\newblock {\em PLoS computational biology}, 9(3):e1002985, jan 2013.

\bibitem{Linkenkaer-Hansen2001}
K~Linkenkaer-Hansen, V~V Nikouline, J~M Palva, and R~J Ilmoniemi.
\newblock {Long-range temporal correlations and scaling behavior in human brain
  oscillations.}
\newblock {\em The Journal of neuroscience : the official journal of the
  Society for Neuroscience}, 21(4):1370--1377, 2001.

\bibitem{Palva2013}
J.~Matias Palva, Alexander Zhigalov, Jonni Hirvonen, Onerva Korhonen, Klaus
  Linkenkaer-Hansen, and Satu Palva.
\newblock {Neuronal long-range temporal correlations and avalanche dynamics are
  correlated with behavioral scaling laws}.
\newblock {\em Proceedings of the National Academy of Sciences},
  110(9):3585--3590, 2013.

\bibitem{Iyer2015}
Kartik~K. Iyer, James~A. Roberts, Lena Hellstr{\"{o}}m-Westas, Sverre
  Wikstr{\"{o}}m, Ingrid {Hansen Pupp}, David Ley, Sampsa Vanhatalo, and
  Michael Breakspear.
\newblock {Cortical burst dynamics predict clinical outcome early in extremely
  preterm infants}.
\newblock {\em Brain}, 138(8):2206--2218, aug 2015.

\bibitem{Tagliazucchi2012}
Enzo Tagliazucchi, Pablo Balenzuela, Daniel Fraiman, and Dante~R. Chialvo.
\newblock {Criticality in large-scale brain fmri dynamics unveiled by a novel
  point process analysis}.
\newblock {\em Frontiers in Physiology}, 3(February):1--12, 2012.

\bibitem{Priesemann2014}
Viola Priesemann, Michael Wibral, Mario Valderrama, Robert Pr{\"{o}}pper,
  Michel {Le Van Quyen}, Theo Geisel, Jochen Triesch, Danko Nikoli{\'{c}}, and
  Matthias H~J Munk.
\newblock {Spike avalanches in vivo suggest a driven, slightly subcritical
  brain state.}
\newblock {\em Frontiers in systems neuroscience}, 8(June):108, jan 2014.

\bibitem{Ribeiro2010}
Tiago~L. Ribeiro, Mauro Copelli, F{\'{a}}bio Caixeta, Hindiael Belchior,
  Dante~R. Chialvo, Miguel a~L Nicolelis, and Sidarta Ribeiro.
\newblock {Spike Avalanches Exhibit Universal Dynamics across the Sleep-Wake
  Cycle}.
\newblock {\em PLoS ONE}, 5(11):e14129, nov 2010.

\bibitem{Bedard2006}
C.~B{\'{e}}dard, H.~Kr{\"{o}}ger, and a.~Destexhe.
\newblock {Does the 1/f frequency scaling of brain signals reflect
  self-organized critical states?}
\newblock {\em Physical Review Letters}, 97(11):1--4, 2006.

\bibitem{Girardi-Schappo2013}
M.~Girardi-Schappo, O.~Kinouchi, and M.~H~R Tragtenberg.
\newblock {Critical avalanches and subsampling in map-based neural networks
  coupled with noisy synapses}.
\newblock {\em Physical Review E}, 88(2):1--5, 2013.

\bibitem{Ribeiro2014}
Tiago~L Ribeiro, Sidarta Ribeiro, Hindiael Belchior, F{\'{a}}bio Caixeta, and
  Mauro Copelli.
\newblock {Undersampled critical branching processes on small-world and random
  networks fail to reproduce the statistics of spike avalanches.}
\newblock {\em PloS one}, 9(4):e94992, jan 2014.

\bibitem{DelPapa2017}
Bruno {Del Papa}, Viola Priesemann, and Jochen Triesch.
\newblock {Criticality meets learning: Criticality signatures in a
  self-organizing recurrent neural network}.
\newblock {\em PLoS ONE}, 12(5):1--22, 2017.

\bibitem{Poil2012}
S.-S. Poil, R.~Hardstone, H.~D. Mansvelder, and K.~Linkenkaer-Hansen.
\newblock {Critical-State Dynamics of Avalanches and Oscillations Jointly
  Emerge from Balanced Excitation/Inhibition in Neuronal Networks}.
\newblock {\em Journal of Neuroscience}, 32(29):9817--9823, 2012.

\bibitem{Touboul2010}
Jonathan Touboul and Alain Destexhe.
\newblock {Can power-law scaling and neuronal avalanches arise from stochastic
  dynamics?}
\newblock {\em PLoS ONE}, 5(2), 2010.

\bibitem{Wilting2019}
Jens Wilting and Viola Priesemann.
\newblock {Between perfectly critical and fully irregular: a reverberating
  model captures and predicts cortical spike propagation}.
\newblock {\em Cereb Cortex}, in press, 2019.

\bibitem{Dahmen2016}
David Dahmen, Markus Diesmann, and Moritz Helias.
\newblock {Distributions of covariances as a window into the operational regime
  of neuronal networks}.
\newblock {\em Preprint at http://arxiv.org/abs/1605.04153}, 2016.

\bibitem{Jaeger2007}
Herbert Jaeger, Wolfgang Maass, and Jose Principe.
\newblock {Special issue on echo state networks and liquid state machines}.
\newblock {\em Neural Networks}, 20(3):287--289, 2007.

\bibitem{Williams-Garcia2014}
Rashid~V. Williams-Garc{\'{i}}a, Mark Moore, John~M. Beggs, and Gerardo Ortiz.
\newblock {Quasicritical brain dynamics on a nonequilibrium Widom line}.
\newblock {\em Physical Review E}, 90(6):062714, dec 2014.

\bibitem{Martinello2017}
Matteo Martinello, Jorge Hidalgo, Amos Maritan, Serena {Di Santo}, Dietmar
  Plenz, and Miguel~A. Mu{\~{n}}oz.
\newblock {Neutral theory and scale-free neural dynamics}.
\newblock {\em Physical Review X}, 7(4):1--11, 2017.

\bibitem{Neto2017}
Joao {Pinheiro Neto} and Viola Priesemann.
\newblock {Coarse sampling bias inference of criticality in neural system}.
\newblock {\em In preparation}.

\bibitem{Dehghani2012}
Nima Dehghani, Nicholas~G. Hatsopoulos, Zach~D. Haga, Rebecca~A. Parker,
  Bradley Greger, Eric Halgren, Sydney~S. Cash, and Alain Destexhe.
\newblock {Avalanche Analysis from Multielectrode Ensemble Recordings in Cat,
  Monkey, and Human Cerebral Cortex during Wakefulness and Sleep}.
\newblock {\em Frontiers in Physiology}, 3(August):1--18, 2012.

\bibitem{Kossio2018}
Felipe Yaroslav~Kalle Kossio, Sven Goedeke, Benjamin van~den Akker, Borja
  Ibarz, and Raoul-Martin Memmesheimer.
\newblock {Growing Critical: Self-Organized Criticality in a Developing Neural
  System}.
\newblock {\em Physical Review Letters}, 121(5):058301, aug 2018.

\bibitem{Brunel2000}
Nicolas Brunel.
\newblock {Dynamics of networks of randomly connected excitatory and inhibitory
  spiking neurons}.
\newblock {\em Journal of Physiology Paris}, 94(5-6):445--463, 2000.

\bibitem{Markram2015}
Henry Markram, Eilif Muller, Srikanth Ramaswamy, Michael~W. Reimann, Marwan
  Abdellah, Carlos~Aguado Sanchez, Anastasia Ailamaki, Lidia Alonso-Nanclares,
  Nicolas Antille, Selim Arsever, Guy Antoine~Atenekeng Kahou, Thomas~K.
  Berger, Ahmet Bilgili, Nenad Buncic, Athanassia Chalimourda, Giuseppe
  Chindemi, Jean~Denis Courcol, Fabien Delalondre, Vincent Delattre, Shaul
  Druckmann, Raphael Dumusc, James Dynes, Stefan Eilemann, Eyal Gal,
  Michael~Emiel Gevaert, Jean~Pierre Ghobril, Albert Gidon, Joe~W. Graham,
  Anirudh Gupta, Valentin Haenel, Etay Hay, Thomas Heinis, Juan~B. Hernando,
  Michael Hines, Lida Kanari, Daniel Keller, John Kenyon, Georges Khazen, Yihwa
  Kim, James~G. King, Zoltan Kisvarday, Pramod Kumbhar, S{\'{e}}bastien
  Lasserre, Jean~Vincent {Le B{\'{e}}}, Bruno~R.C. Magalh{\~{a}}es, Angel
  Merch{\'{a}}n-P{\'{e}}rez, Julie Meystre, Benjamin~Roy Morrice, Jeffrey
  Muller, Alberto Mu{\~{n}}oz-C{\'{e}}spedes, Shruti Muralidhar, Keerthan
  Muthurasa, Daniel Nachbaur, Taylor~H. Newton, Max Nolte, Aleksandr
  Ovcharenko, Juan Palacios, Luis Pastor, Rodrigo Perin, Rajnish Ranjan, Imad
  Riachi, Jos{\'{e}}~Rodrigo Rodr{\'{i}}guez, Juan~Luis Riquelme, Christian
  R{\"{o}}ssert, Konstantinos Sfyrakis, Ying Shi, Julian~C. Shillcock, Gilad
  Silberberg, Ricardo Silva, Farhan Tauheed, Martin Telefont, Maria
  Toledo-Rodriguez, Thomas Tr{\"{a}}nkler, Werner {Van Geit}, Jafet~Villafranca
  D{\'{i}}az, Richard Walker, Yun Wang, Stefano~M. Zaninetta, Javier Defelipe,
  Sean~L. Hill, Idan Segev, and Felix Sch{\"{u}}rmann.
\newblock {Reconstruction and Simulation of Neocortical Microcircuitry}.
\newblock {\em Cell}, 163(2):456--492, 2015.

\bibitem{Palmigiano2017}
Agostina Palmigiano, Theo Geisel, Fred Wolf, and Demian Battaglia.
\newblock {Flexible information routing by transient synchrony}.
\newblock {\em Nature Neuroscience}, 20(7):1014--1022, may 2017.

\bibitem{DiSanto2018}
Serena di~Santo, Pablo Villegas, Raffaella Burioni, and Miguel~A. Mu{\~{n}}oz.
\newblock {Landau–Ginzburg theory of cortex dynamics: Scale-free avalanches
  emerge at the edge of synchronization}.
\newblock {\em Proceedings of the National Academy of Sciences},
  115(7):E1356--E1365, feb 2018.

\bibitem{Mora2011}
Thierry Mora and William Bialek.
\newblock {Are Biological Systems Poised at Criticality?}
\newblock {\em Journal of Statistical Physics}, 144(2):268--302, 2011.

\bibitem{Tkacik2013}
Ga{\v{s}}per Tka{\v{c}}ik, Olivier Marre, Thierry Mora, Dario Amodei,
  Michael~J. Berry, and William Bialek.
\newblock {The simplest maximum entropy model for collective behavior in a
  neural network}.
\newblock {\em Journal of Statistical Mechanics: Theory and Experiment},
  2013(3), 2013.

\bibitem{Tkacik2015}
Ga{\v{s}}per Tka{\v{c}}ik, Thierry Mora, Olivier Marre, Dario Amodei,
  Stephanie~E. Palmer, Michael~J. Berry, and William Bialek.
\newblock {Thermodynamics and signatures of criticality in a network of
  neurons}.
\newblock {\em Proceedings of the National Academy of Sciences},
  112(37):11508--11513, sep 2015.

\bibitem{Humplik2017}
Jan Humplik and Ga{\v{s}}per Tka{\v{c}}ik.
\newblock {Probabilistic models for neural populations that naturally capture
  global coupling and criticality}.
\newblock {\em PLoS Computational Biology}, 13(9):1--26, 2017.

\bibitem{Savin2017}
Cristina Savin and Ga{\v{s}}per Tka{\v{c}}ik.
\newblock {Maximum entropy models as a tool for building precise neural
  controls}.
\newblock {\em Current Opinion in Neurobiology}, 46:120--126, 2017.

\bibitem{Nonnenmacher2017}
Marcel Nonnenmacher, Christian Behrens, Philipp Berens, Matthias Bethge, and
  Jakob~H. Macke.
\newblock {Signatures of criticality arise from random subsampling in simple
  population models}.
\newblock {\em PLoS Computational Biology}, 13(10):1--23, 2017.

\bibitem{Murray2014a}
John~D Murray, Alberto Bernacchia, David~J Freedman, Ranulfo Romo, Jonathan~D
  Wallis, Xinying Cai, Camillo Padoa-Schioppa, Tatiana Pasternak, Hyojung Seo,
  Daeyeol Lee, and Xiao-Jing Wang.
\newblock {A hierarchy of intrinsic timescales across primate cortex.}
\newblock {\em Nature neuroscience}, 17(12):1661--3, 2014.

\bibitem{Hasson2008}
U.~Hasson, E.~Yang, I.~Vallines, D.~J. Heeger, and N.~Rubin.
\newblock {A Hierarchy of Temporal Receptive Windows in Human Cortex}.
\newblock {\em Journal of Neuroscience}, 28(10):2539--2550, 2008.

\bibitem{Honey2012}
Christopher~J. Honey, Thomas Thesen, Tobias~H. Donner, Lauren~J. Silbert,
  Chad~E. Carlson, Orrin Devinsky, Werner~K. Doyle, Nava Rubin, David~J.
  Heeger, and Uri Hasson.
\newblock {Slow Cortical Dynamics and the Accumulation of Information over Long
  Timescales}.
\newblock {\em Neuron}, 76(2):423--434, 2012.

\bibitem{Chen2015}
Janice Chen, Uri Hasson, and Christopher~J. Honey.
\newblock {Processing Timescales as an Organizing Principle for Primate
  Cortex}.
\newblock {\em Neuron}, 88(2):244--246, 2015.

\bibitem{Chaudhuri2015}
Rishidev Chaudhuri, Kenneth Knoblauch, Marie~Alice Gariel, Henry Kennedy, and
  Xiao~Jing Wang.
\newblock {A Large-Scale Circuit Mechanism for Hierarchical Dynamical
  Processing in the Primate Cortex}.
\newblock {\em Neuron}, 88(2):419--431, 2015.

\bibitem{Meisel2017a}
Christian Meisel, Andreas Klaus, Vladyslav~V. Vyazovskiy, and Dietmar Plenz.
\newblock {The Interplay between Long- and Short-Range Temporal Correlations
  Shapes Cortex Dynamics across Vigilance States}.
\newblock {\em The Journal of Neuroscience}, 37(42):10114--10124, oct 2017.

\bibitem{Meisel2017b}
Christian Meisel, Kimberlyn Bailey, Peter Achermann, and Dietmar Plenz.
\newblock {Decline of long-range temporal correlations in the human brain
  during sustained wakefulness}.
\newblock {\em Scientific Reports}, 7(1):11825, dec 2017.

\bibitem{Tagliazucchi2016}
Enzo Tagliazucchi, Dante~R. Chialvo, Michael Siniatchkin, Enrico Amico,
  Jean-Francois Brichant, Vincent Bonhomme, Quentin Noirhomme, Helmut Laufs,
  and Steven Laureys.
\newblock {Large-scale signatures of unconsciousness are consistent with a
  departure from critical dynamics}.
\newblock {\em Journal of The Royal Society Interface}, 13(114):20151027, 2016.

\bibitem{Fagerholm2016}
Erik~D. Fagerholm, Gregory Scott, Woodrow~L. Shew, Chenchen Song, Robert Leech,
  Thomas Kn{\"{o}}pfel, and David~J. Sharp.
\newblock {Cortical Entropy, Mutual Information and Scale-Free Dynamics in
  Waking Mice}.
\newblock {\em Cerebral Cortex}, 26(10):3945--3952, 2016.

\bibitem{Fagerholm2018}
Erik~D. Fagerholm, Martin Dinov, Thomas Kn{\"{o}}pfel, and Robert Leech.
\newblock {The characteristic patterns of neuronal avalanches in mice under
  anesthesia and at rest: An investigation using constrained artificial neural
  networks}.
\newblock {\em PLoS ONE}, 13(5):1--19, 2018.

\bibitem{Fekete2018}
Tomer Fekete, David~B. Omer, Kazunori O'Hashi, Amiram Grinvald, Cees van
  Leeuwen, and Oren Shriki.
\newblock {Critical dynamics, anesthesia and information integration: Lessons
  from multi-scale criticality analysis of voltage imaging data}.
\newblock {\em NeuroImage}, 183(November 2017):919--933, 2018.

\bibitem{Lee2019}
Heonsoo Lee, Daniel Golkowski, Denis Jordan, Sebastian Berger, R{\"{u}}diger
  Ilg, Joseph Lee, George~A. Mashour, Uncheol Lee, Michael~S. Avidan, Stefanie
  Blain-moraes, Goodarz Golmirzaie, Randall Hardie, Rosemary Hogg, Ellen Janke,
  Max~B. Kelz, Kaitlyn Maier, George~A. Mashour, Hannah Maybrier, Andrew
  Mckinstry-wu, and Maxwell Muench.
\newblock {Relationship of critical dynamics, functional connectivity, and
  states of consciousness in large-scale human brain networks}.
\newblock {\em NeuroImage}, 188(December 2018):228--238, 2019.

\bibitem{Fagerholm2015}
E.~D. Fagerholm, R.~Lorenz, G.~Scott, M.~Dinov, P.~J. Hellyer, N.~Mirzaei,
  C.~Leeson, D.~W. Carmichael, D.~J. Sharp, W.~L. Shew, and R.~Leech.
\newblock {Cascades and Cognitive State: Focused Attention Incurs Subcritical
  Dynamics}.
\newblock {\em Journal of Neuroscience}, 35(11):4626--4634, 2015.

\bibitem{Arviv2015}
O.~Arviv, A.~Goldstein, and O.~Shriki.
\newblock {Near-Critical Dynamics in Stimulus-Evoked Activity of the Human
  Brain and Its Relation to Spontaneous Resting-State Activity}.
\newblock {\em Journal of Neuroscience}, 35(41):13927--13942, 2015.

\bibitem{Barnett2013}
Lionel Barnett, Joseph~T. Lizier, Michael Harr{\'{e}}, Anil~K. Seth, and Terry
  Bossomaier.
\newblock {Information flow in a kinetic ising model peaks in the disordered
  phase}.
\newblock {\em Physical Review Letters}, 111(17):1--4, 2013.

\end{thebibliography}

\end{scriptsize}

\end{document}